\documentclass[10pt]{iopart}

\usepackage{iopams}
\usepackage{graphicx}
\usepackage{bm}
\usepackage{natbib}

\newcommand{\text}[1]{\mathrm{#1}}
\newcommand{\eqref}[1]{(\ref{#1})}

\newcommand{\const}{\ensuremath{\text{const}}}
\newcommand{\abs}[1]{\ensuremath{\left| #1\right|}}
\newcommand{\E}[1]{\ensuremath{\hspace{1pt}\mathrm{e}^{#1}}}

\newcommand{\be}{\begin{equation}}
\newcommand{\ee}{\end{equation}}
\renewcommand{\bs}{\numparts}
\renewcommand{\es}{\endnumparts}

\newcommand{\ex}{\ensuremath{\hat{\boldsymbol{e}}_x}}

\newcommand{\se}{\ensuremath{_\perp}}

\newcommand{\Ph}{\ensuremath{\Phi}}

\newcommand{\La}{\ensuremath{\Lambda}}
\newcommand{\Ps}{\ensuremath{\Psi}}

\newcommand{\uint}{\ensuremath{\int_{-\infty}^\infty}}

\newcommand{\f}[1]{\ensuremath{\boldsymbol{#1}}}

\newcommand{\pd}[2][]{\ensuremath{\frac{\partial #1}{\partial #2}}}
\newcommand{\dd}[2][]{\ensuremath{\frac{\mathrm{d} #1}{\mathrm{d} #2}}}
\newcommand{\df}{\ensuremath{\mathrm{d}}}

\begin{document}

\title[Non-linear Weibel-type Soliton Modes]
{Non-linear Weibel-type Soliton Modes}

\author{R.\,C.~Tautz$^1$, I.~Lerche$^2$}

\eads{\mailto{rct@gmx.eu}, \mailto{lercheian@yahoo.com}}

\vspace{5pt}
\address{$^1$Department of Astronomy and Astrophysics, Berlin Institute of Technology, Hardenbergstra\ss{}e~36, D--10623 Berlin, Germany\\[2pt]
$^2$Institut f\"ur Geowissenschaften, Naturwissenschaftliche Fakult\"at III,\\
Martin-Luther-Universit\"at Halle, D-06099 Halle, Germany}

\date{\today}

\begin{abstract}
Discussion is given of non-linear soliton behavior including coupling between electrostatic and electromagnetic potentials for non-relativistic, weakly relativistic, and fully relativistic plasmas. For plasma distribution functions that are independent of the canonical momenta perpendicular to the soliton spatial structure direction there are, in fact, no soliton behaviors allowed because transverse currents are zero. Dependence on the transverse canonical momenta is necessary. When such is the case, it is shown that the presence or absence of a soliton is intimately connected to the functional form assumed for the particle distribution functions. Except for simple situations, the coupled non-linear equations for the electrostatic and electromagnetic potentials would seem to require numerical solution procedures. Examples are given to illustrate all of these points for non-relativistic, weakly relativistic, and fully relativistic plasmas.
\end{abstract}

\hspace{2.5cm}{\it Keywords\/} relativistic plasmas, transverse waves, solitons

\section{Introduction}

Since its discovery in 1959, the Weibel instability \citep{wei59:wei,fri59:wei} has been, and continues to be, a subject of intense research. Part of this focus is due to the fact that the Weibel instability excites aperiodic fluctuations, i.\,e., purely growing waves that do not propagate out of the system. Therefore, the Weibel instability has been considered to be responsible for the creation of seed magnetic fields in the early universe \citep{sch03:cos,sak04:mag,sch05:ori}, which act as a progenitor to the large-scale magnetic field that we observe today in all (spiral) galaxies \citep{bec96:mag}. Likewise, Weibel fluctuations can provide the necessary dissipation of bulk velocity that leads to the formation of shock waves \citep{tau06:grb}. In highly relativistic jets that occur during events such as GRBs (Gamma-Ray Bursts) or in extreme objects such as AGNs (Active Galactic Nuclei), plasma instabilities \citep{sch02:agn} and particularly the Weibel instability are, thus, ultimately responsible for the radiation observed at Earth.

Weibel modes and their asssociated non-linear structures \citep{byc03:wei} also play a role in the radiative cooling of relativistic particles in blazar and gamma-ray burst sources \citep{sch07:coo}. Furthermore, Weibel modes can also be excited in quantum plasmas \citep{haa08:qua,haa08:mac}, thus generalizing the classical Weibel instability equation; here, a connection has been made to (dark) soliton waves \citep{shu06:las}. It is also worth noticing that there exist experimental verifications of Weibel \citep{wei01:exp} and soliton \citep{bor02:exp} modes in laser plasmas, thus emphasizing the broad applicability of the underlying mechanism, which converts the free energy from anisotropic distributions into magnetic field energy. In analytical studies, such soliton modes have been used to create magnetic turbulence \citep{kin73:tur,wei77:pla}.

The underlying analytical investigation of the non-linear aspects of the classic Weibel instability made use of the fact that the classic Weibel instability excites only transverse, electromagnetic fluctuations \citep{usr06:nli}. For asymmetric distributions, it was shown that the range of unstable wavenumbers is reduced to one single unstable wave mode, which reminds one of solitary structures that are based on single wavenumbers, too. For the case of transverse electromagnetic modes, it was shown that, depending on the exact form of the distribution function, spatially limited structures are produced (solitons).

From the radiation pattern of particles scattered in soliton modes \citep{usr09:rad}, it is known that there are many similarities to synchrotron radiation. Motivated by the fact that the energy output of the particles is mostly referred to as synchrotron radiation and inverse Compton scattering, it is a necessity to take into account other radiation processes, also because synchrotron radiation requires the presence of a background magnetic field. Until now, the origin and themechanism for such a field have been discussed intensely \citep{med99:grb,med07:wei,med07:jit}. The train of thoughts is the following: (i) in reality, particle velocity distributions should almost always be somewhat asymmetric; (ii) in this case, unstable plasma modes are monochromatic, as has been demonstrated analytically \citep{tau06:is1}; (iii) such isolated structures can lead to soliton modes as has been shown for purely electromagnetic Weibel modes \citep{usr06:nli}; (iv) in relativistic outflows such as for GRBs and AGN jets, all kinds of plasma instabilities are expected to arise. Hence, it is most important to discuss the radiation pattern for such scenarios that is generated by particles being scattered in such magnetic structures. In principle, the method of obtaining the differential frequency spectrum is similar to that for synchrotron radiation \citep{usr09:rad}. The basic difference is that for synchrotron radiation the magnetic field is considered to be constant and the electron moves in circles perpendicular to the magnetic field, while in the case of the soliton the electrons move mostly linearly and are deflected via the Lorentz force. Thus, the radiation is not produced by acceleration through a constant background field but, instead, is caused by an interaction of the electrons with highly varying magnetic and electric fields.

Inspired by the procedure shown in \citet{usr06:nli}, a number of subsequent, more detailed, investigations revealed the exotic properties of the (linear) Weibel instability that are unfolded in the case of totally asymmetric distribution functions \citep{tau06:is1,tau06:is2,tau07:evi,tau07:tun}. For such distributions, the electrostatic and electromagnetic wave modes are coupled, and it was shown that any unstable Weibel mode must be isolated, i.\,e., restricted to a single discrete wavenumber. Specific examples for the distribution function illustrated that isolated Weibel modes are excited. Even if one allows for a small real part of the frequency, the isolated Weibel modes persist \citep{tau07:wea}. Such weakly propagating unstable modes are important for oblique wave propagation, because for such cases the growth rate of unstable waves is maximal \citep{bre04:bpi,bre05:bpi}. It is then appropriate to ask how non-linear soliton modes are influenced when one includes the coupling between electrostatic and electromagnetic potentials. The purpose of this paper is to explore that question.

The paper is organized as follows. In Sec.~\ref{tech}, the Vlasov equation is transformed and the non-linear wave equations are derived. In Sec.~\ref{nonrel}, the non-relativistic soliton behavior is derived and examples are given for two basic types of distribution functions. In Sec.~\ref{rel}, the weakly and the fully relativistic behaviors are derived, which requires approximations as regards the transformation of the volume element in momentum space. Furthermore, in Secs.~\ref{nonrel} and \ref{rel}, solutions are derived for two limiting cases regarding the relative values of the electrostatic and vector potentials. In Sec.~\ref{summ}, the results are summarized and discussed. Detailed explanations of the integral transformations in the cases of non-relativistic and relativistic plasmas are given in \ref{nr_transf} and \ref{r_transf}, respectively.

\section{The Relativistic Vlasov Equation}\label{tech}

Throughout the derivation of the non-linear wave equation and the instability conditions, the vector potential, $\f A=(A_x,A_y,A_z)$, and the scalar potential, $\Ph$, will be used, with
\bs\begin{eqnarray}
\f E&=&-\frac{1}{c}\pd[\f A]t-\f\nabla\Ph\\
\f B&=&\f\nabla\times\f A.
\end{eqnarray}\es
For a soliton wave, observed to be moving at a constant velocity $\beta c$, the easiest way to handle the non-linear equations is to transform to a reference frame moving with the soliton so that, in such a reference frame, the potentials $\f A$ and $\Ph$ are functions solely of the spatial coordinate, $x$, and are independent of time. The electric and magnetic fields are then given by
\bs
\begin{eqnarray}
\f E&=&-\Ph'\ex\\
\f B&=&(0,-A'_z,A'_y),
\end{eqnarray}
\es
where primes denote differentiation with respect to $x$. The Vlasov equation can then be expressed as\footnote{While our notation differs from that used in classical relativistic mechanics, it is traditional to the field of plasma physics.}
\be\label{eq:vlas1}
\fl
0=v_x\,\pd[f]x+\frac{e}{c}\left[\pd[f]{p_x}\left(-c\Ph'+v_yA'_y+v_zA'_z\right)-\pd[f]{p_y}\,A'_yv_x-\pd[f]{p_z}\,A'_zv_x\right].
\ee
The characteristics of Eq.~\eqref{eq:vlas1} with respect to the $y$ and $z$ coordinates are given through $\df p_{(y,z)}/\df A_{(y,z)}=-e/c$ \citep[compare with][]{usr06:nli}. For the solution of the characteristic equations introduce $\varpi_{(y,z)}$ through
\be
\varpi_{(y,z)}=p_{(y,z)}+\frac{e}{c}\,A_{(y,z)}.
\ee
Then, Eq.~\eqref{eq:vlas1} simplifies considerably, because $\varpi_{(y,z)}$ are constants and thus the partial differentiations of the distribution function $f$ with respect to $\varpi_y$ and $\varpi_z$ vanish, yielding
\be\label{eq:vlas2}
\fl
0=\frac{p_x}{m\gamma}\,\pd[f]x+\frac{e}{c}\,\pd[f]{p_x}\left[-c\Ph'+\frac{1}{m\gamma}\left(\varpi_y-\frac{e}{c}\,A_y\right)A'_y+\frac{1}{m\gamma}\left(\varpi_z-\frac{e}{c}\,A_z\right)A'_z\right],
\ee
with $f=f(x,p_x,\varpi_y,\varpi_z)$ and where
\be
\gamma^2=1+\frac{1}{(mc)^2}\left[p_x^2+\left(\varpi_y-\frac{e}{c}\,A_y\right)^2+\left(\varpi_z-\frac{e}{c}\,A_z\right)^2\right].
\ee

It is instructive to first consider the non-relativistic limit [set $\gamma=1$ in Eq.~\eqref{eq:vlas2}] because this limit allows exact solution of the Vlasov equation for arbitrary $\Ph,A_y$, and $A_z$. The non-linear coupling of the electrostatic and electromagnetic potentials in the Maxwell equations can then be investigated simply in the weak coupling limit, where ``weak'' will be defined later.

Thereafter one can consider the relativistic particle behavior where, as will be shown, coupling of the electrostatic and electromagnetic potentials is considerably more involved, although the basic procedure from the non-relativistic situation can be used, \emph{mutatis mutandis}.

\section{The Non-Relativistic Limit}\label{nonrel}

In the case where one deals solely with non-relativistic particles one can set $\gamma=1$ in Eq.~\eqref{eq:vlas2}. Then one has
\be\label{eq:vlas3}
\fl
0=\frac{p_x}{m}\,\pd[f]x+e\,\pd[f]{p_x}\left[-\Ph'+\frac{1}{mc}\left(\varpi_y-\frac{e}{c}\,A_y\right)A'_y+\frac{1}{mc}\left(\varpi_z-\frac{e}{c}\,A_z\right)A'_z\right].
\ee
Set $p_x^2=u$ in Eq.~\eqref{eq:vlas3} to obtain
\be\label{eq:vlas4}
\fl
0=\pd[f]x+2em\,\pd[f]u\left[-\Ph'+\frac{1}{mc}\left(\varpi_y-\frac{e}{c}\,A_y\right)A'_y+\frac{1}{mc}\left(\varpi_z-\frac{e}{c}\,A_z\right)A'_z\right].
\ee
With
\be
w=u+2em\Ph+\left(\varpi_y-\frac{e}{c}\,A_y\right)^2+\left(\varpi_z-\frac{e}{c}\,A_z\right)^2,
\ee
Eq.~\eqref{eq:vlas4} provides the general exact solution $f=f(w,\varpi_y,\varpi_z)$ by direct substitution.\footnote{Consider Eq.~\eqref{eq:vlas4} in the form $\partial f/\partial x-(\df g/\df x)(\partial f/\partial u)=0$. Following the theory of linear partial differential equations, set $g(x)$ as a new variable. Then $\partial f/\partial g-\partial f/\partial u=0$, which has the general solution $f=f(g+u)$. With $w=g+u$, this is precisely as given.}

Consider then Maxwell's equations in the limit of non-relativistic particles. For a plasma with different constituents---denoted by the index $\alpha$ with, e.\,g., $\alpha=e$ for electrons and $\alpha=p$ for protons---one has
\bs
\begin{eqnarray}
\Ph''&=&-4\pi\sum_\alpha e_\alpha\int\df^3p\;f_\alpha\\
A''_{(y,z)}&=&-4\pi\sum_\alpha\frac{e_\alpha}{m_\alpha c}\int\df^3p\;p_{(y,z)}f_\alpha
\end{eqnarray}
\es
In terms of $w,\varpi_y$, and $\varpi_z$ the Maxwell equations take the form (see \ref{nr_transf})
\bs
\be\label{eq:Ph1}
\fl
\Ph''=8\pi\sum_\alpha e_\alpha\int_0^\infty\df\xi\uint\df\varpi_y\uint\df\varpi_z\;\sqrt\xi\,\pd\xi\,f_\alpha\!\left(w_\star+\xi,\varpi_y,\varpi_z\right)
\ee
and
\begin{eqnarray}
\f A''\se&=&8\pi\sum_\alpha\frac{e_\alpha}{m_\alpha c}\int_0^\infty\df\xi\uint\df\varpi_y\uint\df\varpi_z\;\sqrt\xi\nonumber\\
&\times&\left(\f\varpi\se-\frac{e_\alpha}{c}\,\f A\se\right)\pd\xi\,f\!\left(w_\star+\xi,\varpi_y,\varpi_z\right) \label{eq:A1},
\end{eqnarray}
\es
with $\f A\se=(0,A_y,A_z)$, $\f\varpi\se=(0,\varpi_y,\varpi_z)$ and $w\equiv w_\star+\xi$, where
\be
w_\star=2e_\alpha m_\alpha\Ph+\left(\f\varpi\se-\frac{e_\alpha}{c}\,\f A\se\right)^2.
\ee

Solution of the coupled electrostatic potential, Eq.~\eqref{eq:Ph1}, and the electromagnetic potential, Eq.~\eqref{eq:A1}, depends sensitively on the assignment prescribed for the particle distribution function $f(w,\varpi_y,\varpi_z)$. Note that if $f(w,\varpi_y,\varpi_z)$ is a function solely of $w$ and not of $\varpi_y,\varpi_z$ \emph{explicitly}, then $\f A\se=0$. Then only $\Ph$ is eligible for a soliton structure. To illustrate this point more closely, consider two cases, where $f(w,\varpi_y,\varpi_z)$ takes on the functional forms
\bs
\be
f=f_{a,0}\exp\!\left(-\frac{w}{w_\alpha}\right) \label{eq:fs1}
\ee
or
\be
f_\alpha=f_{a,0}\exp\!\left(-\frac{w}{w_\alpha}\right)\exp\!\left(-\frac{\varpi_y^2+\varpi_z^2}{\varpi_\alpha^2}\right), \label{eq:fs2}
\ee
\es
with $f_{a,0}$, $w_\alpha$, and $\varpi_\alpha$ constants.

Direct integration of the right-hand sides at Eqs.~\eqref{eq:Ph1} and \eqref{eq:A1} is possible with the expressions from Eqs.~\eqref{eq:fs1} and \eqref{eq:fs2}. The results are in the case of Eq.~\eqref{eq:fs1}
\bs
\begin{eqnarray}
\Ph''&=&-4\pi^{5/2}\sum_\alpha f_{a,0}e_\alpha w_\alpha^{3/2}\exp\!\left(-\frac{2e_\alpha m_\alpha}{w_\alpha}\,\Ph\right) \label{eq:case1a}\\
\f A''\se&=&0 \label{eq:case1b}
\end{eqnarray}
\es
and in the case of Eq.~\eqref{eq:fs2}
\bs
\be\label{eq:case2a}
\fl
\Ph''=-4\pi^{5/2}\sum_\alpha f_{a,0}\frac{e_\alpha w_\alpha^{3/2}\varpi_\alpha^2}{\varpi_\alpha^2+w_\alpha}\,\exp\!\left[-\frac{2e_\alpha m_\alpha}{w_\alpha}\,\Ph-\left(\frac{e_\alpha}{c}\right)^2\frac{A_y^2+A_z^2}{w_\alpha+\varpi_\alpha^2}\right]
\ee
together with
\begin{eqnarray}
A''_j&=&4\pi^{5/2}\sum_\alpha f_{a,0}\left(\frac{e_\alpha}{c}\right)^2\frac{w_\alpha^2}{\left(w_\alpha+\varpi_\alpha^2\right)^2}\,A_j\nonumber\\
&\times&\exp\!\left[-\frac{2e_\alpha m_\alpha}{w_\alpha}\Ph-\left(\frac{e_\alpha}{c}\right)^2\frac{A_y^2+A_z^2}{w_\alpha+\varpi_\alpha^2}\right] \label{eq:case2b}
\end{eqnarray}
\es
with $j\in\{y,z\}$. Note that as $\varpi_\alpha\to\infty$ Eqs.~\eqref{eq:case2a} and \eqref{eq:case2b} reduce to Eqs.~\eqref{eq:case1a} and \eqref{eq:case1b}, respectively, as required. The determination of a soliton structure in both cases can then be readily given for different plasma conditions.

For instance an electron-positron plasma with identical plasma characteristics (i.\,e., $f_{a,0}$ the same for both species) allows one to write Eq.~\eqref{eq:case1a} in the form
\be
\Ph''=4\pi^{5/2}f_\alpha e_\alpha w_\alpha^{3/2}\left[\exp\!\left(\frac{2e_\alpha m_\alpha}{w_\alpha}\,\Ph\right)-\exp\!\left(-\frac{2e_\alpha m_\alpha}{w_\alpha}\,\Ph\right)\right],
\ee
which integrates once directly to yield
\be\label{eq:case1a2}
\left(\Ph'\right)^2=\frac{4\pi^{5/2}w_\alpha^{5/2}}{m_\alpha}\,f_\alpha\cosh\!\left(\frac{2m_\alpha e_\alpha}{w_\alpha}\,\Ph\right)+\const.
\ee
With $\Ps=2m_\alpha e_\alpha\Ph/w_\alpha$ one has
\be\label{eq:Psp1}
\left(\Ps'\right)^2=\frac{4\pi^{5/2}w_\alpha^{5/2}}{m_\alpha}\left(\frac{4m_\alpha^2e_\alpha^2}{w_\alpha^2}\right)f_\alpha\cosh\Ps+\const.
\ee

Set the constant to a negative value, i.\,e., $-\cosh\Ps_\star$ (a positive constant would automatically keep $\Ps'$ growing and so cannot represent a soliton), yielding
\be\label{eq:Psp2}
\left(\Ps'\right)^2=16\pi^{5/2}w_\alpha^{1/2}f_\alpha m_\alpha e_\alpha^2\left(\cosh\Ps-\cosh\Ps_\star\right).
\ee
Then if \Ps\ exceeds $\Ps_\star$ it does so thereafter for all coordinates and so does not represent a soliton. If \Ps\ is less than $\Ps_\star$ then Eq.~\eqref{eq:Psp2} is not valid because it would yield $(\Ps')^2$ negative. (For negative \Ps, the argument can be reversed.) Hence the only solution is a spatially unbounded potential \Ps\ which does not correspond to a soliton.

Such is in the line with the small amplitude limit that provides a dispersion relation of the form $\omega^2=\omega_p^2+k^2v_{\text{th}}^2$ which, for $\omega=0$, indicates $k=\pm i\omega_p/v_{\text{th}}$ and so represents spatially unbounded growing or decaying modes. Such unbounded structures do not represent soliton modes and so are to be discarded. The only solution left for Eq.~\eqref{eq:case1a2} is then $\Ph=0$.

\begin{figure}[t]
\centering
\includegraphics[width=\linewidth]{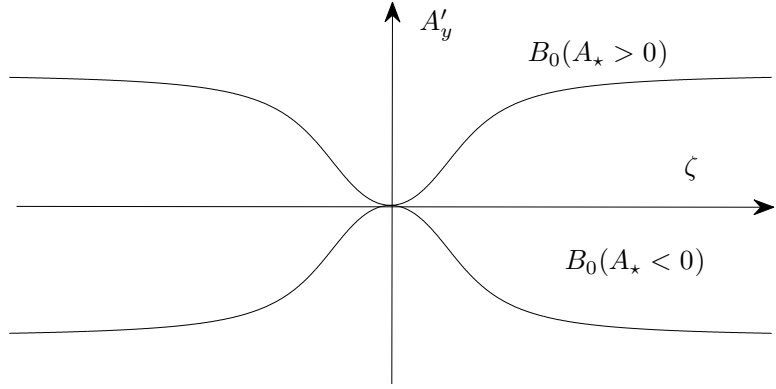}
\caption{The behavior of $A'_y$ as a function of the normalized variable, $\zeta$, for positive and negative values of $B_0$ at $A=A_\star$, respectively.}
\label{ab:f1}
\end{figure}

In the case of Eq.~\eqref{eq:fs2} there is a common factor in Eq.~\eqref{eq:case2a} of
\begin{equation*}
\exp\!\left[-\left(\frac{e_\alpha}{c}\right)^2\frac{A_y^2+A_z^2}{w_\alpha+\varpi_\alpha^2}\right]
\end{equation*}
so that for the electron-positron plasma one again has $\Ph''>0$ everywhere and so no electrostatic soliton is available. Here, ``common'' means that the factor does not depend on summation over species and so can be brought outside the summation. Thus $\Ph=0$ is the only acceptable solution of Eq.~\eqref{eq:case2a}. Equation~\eqref{eq:case2b} takes on the generic form
\be
A''_j=\beta\,A_j\exp\!\left[-\alpha\left(A_y^2+A_z^2\right)\right]
\ee
with $\alpha,\beta>0$ and $j\in\{y,z\}$. Then either $A''_y/A_y=A''_z/A_z$ or one of the components $A_y$ or $A_z$ is zero. Consider that $A_z=0$. Then
\be
A''_y=\beta\,A_y\exp\!\left[-\alpha A_y^2\right],
\ee
which integrates once to provide
\be\label{eq:tmp1}
\left(A'_y\right)^2=\frac{\beta}{\alpha}\left[\const-\exp\!\left(-\alpha A_y^2\right)\right].
\ee

\begin{figure}[t]
\centering
\includegraphics[width=\linewidth]{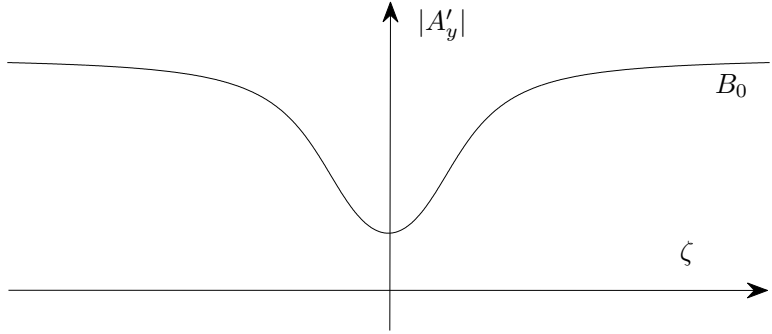}
\caption{The behavior of the absolute value of $A'_y$ as a function of the normalized variable, $\zeta$.}
\label{ab:f2}
\end{figure}

Let $A_y$ have an extremum on $A_y=A_\star$. Then write Eq.~\eqref{eq:tmp1} in the form
\be\label{eq:Ayp}
\left(A'_y\right)^2=B_0^2\left\{1-\exp\!\left[-\alpha\left(A_y^2-A_\star^2\right)\right]\right\}
\ee
with
\begin{equation*}
\exp\!\left(-\alpha A_\star^2\right)=\alpha B_0^2/\beta
\end{equation*}
so that $A'_y=0$ on $A_y=A_\star$. Eq.~\eqref{eq:Ayp} shows that $A'_y$ has the structure given in Fig.~1 with the two solutions only just touching on $A'_y=0$ but not crossing because
\begin{equation*}
A''_y\gtrless0 \text{\;on\;} A_y=A_\star,\qquad \text{for\;} A_\star\gtrless0,
\end{equation*}
representing \emph{both} a background magnetic field $B_0$ together with a soliton pulse superposed. The absolute value, $\abs{A'_y}$, is illustrated in Fig.~\ref{ab:f2}. Although Eq.~\eqref{eq:Ayp} is not analytically resolvable, the structure of the solution is indeed seen by considering turning points and roots.

The point, then, is that exact non-linear solutions to the coupled Maxwell-Vlasov equations can be arranged to provide a rich variety of structural behaviors depending on the specific choices made for the distribution functions for the particles. In order that there be an electrostatic soliton component in Eq.~\eqref{eq:Ph1} and \eqref{eq:A1} one requires that the precise symmetry invoked between electrons and positrons be broken. Such a break can be arranged in one of two ways: either one chooses different distribution functions for the positive and negative charged components or one chooses different particle masses for the positive ions with respect to the negatively charged particles.

While it was shown that one class of solution is that when one of the components $A_j$ (with $j\in\{y,z\}$) is zero, that does not rule out other classes of solution where $A_j$ is non-zero. In addition one cannot provide a general statemenet on the field direction relative to the background field because each choice of distribution function must be investigated for its effects on the non-linear system. It also by no means follows that a plasma with no background field cannot posses solitons of electrostatic or electromagnetic or coupled properties---again, each distribution function chosen for each species provides its own properties- which was part of the aim of the special illustrations chosen.

For example, if one were to treat with finite electrons and infinitely massive ions then, with the distribution function from Eq.~\eqref{eq:fs1} and \eqref{eq:fs2}, one obtains
\bs
\be\label{eq:massPh}
\fl
\Ph''=4\pi^{5/2}f_{e,0}\,\frac{ew_e^{3/2}\varpi_e^2}{w_e+\varpi_e^2}\,\exp\!\left[\frac{2em}{w_e}\,\Ph-\left(\frac{e}{c}\right)^2\frac{A_y^2+A_z^2}{w_e+\varpi_e^2}\right]-4\pi en,
\ee
where $n$ is the ion number density and
\be\label{eq:massA}
\fl
A''_j=4\pi^{5/2}f_{e,0}\left(\frac{e}{c}\right)^2\frac{w_e^2}{\left(w_e+\varpi_e^2\right)^2}\,A_j\,\exp\!\left[\frac{2em}{w_e}\,\Ph-\left(\frac{e}{c}\right)^2\frac{A_y^2+A_z^2}{w_e+\varpi_e^2}\right]
\ee
\es
with $j\in\{y,z\}$.

Note that one class of solutions is when $A_j=0$ and when such is the case Eq.~\eqref{eq:massPh} has the structural form
\be\label{eq:massPs}
\Ps''=b\,\exp(\Ps)-gn,\qquad b>0,g>0
\ee
Eq.~\eqref{eq:massPs} integrates once immediately to yield
\be\label{eq:yld}
\left(\Ps'\right)^2=2b\E\Ps-2gn\Ps+\La,
\ee
where \La\ is a constant. Consider now in detail the structure of Eq.~\eqref{eq:yld}. Write $\nu=\Ps-\La/(2gn)$ where one has
\be
\left(\nu'\right)^2=B\E\nu-2gn\nu
\ee
with $B=2b\exp[\La/(2gn)]$. Further simplification occurs when one sets $x=(2gn)^{-1/2}\zeta$ and $B/(2gn)=R^2$ when one has
\be\label{eq:nuzeta}
\left(\dd[\nu]\zeta\right)^2=R^2\E\nu-\nu.
\ee
Note that \La, the constant of integration, appears through the relationship between $B$ and $R$. Only for specific ranges of \La\ can one expect to obtain a soliton behavior as we now show. On $\nu=0$, Eq.~\eqref{eq:nuzeta} yields
\begin{equation*}
\dd[\nu]\zeta=\pm R
\end{equation*}
so that two branching structures exist. If a soliton is to exist then it is necessary that $\df\nu/\df\zeta=0$ somewhere, implying a value $\nu_\star$ such that
\be\label{eq:nustar}
R^2\E{\nu_\star}=\nu_\star
\ee
Solutions to Eq.~\eqref{eq:nustar} exist only when $R^2<\E{-1}$ and, under that condition, there are two positive values of $\nu_\star$, namely, $\nu_{\text L}$ and $\nu_{\text U}$, where $\nu_{\text U}>\nu_{\text L}$ without loss of generality. Now Eq.~\eqref{eq:nuzeta} requires $(\df\nu/\df\zeta)^2\geqslant0$ everywhere. One therefore has to distinguish three cases:

\begin{itemize}
\item In $\nu_{\text L}<\nu<\nu_{\text U}$ one has $R^2\E\nu<\nu$ so that Eq.~\eqref{eq:nuzeta} cannot be satisfied. Hence the potential regimes for solitons are either $\nu>\nu_{\text U}$ or $\nu<\nu_{\text L}$.

\item For $\nu>\nu_{\text U}$ one has $(\df\nu/\df\zeta)^2>0$ so that no bounded structure is possible.

\item Equally for $\nu<\nu_{\text L}$ one has the same argument with the added problem that $(\df\nu/\df\zeta)^2$ grows indefinitely in $\nu<0$. Thus no electrostatic soliton is possible.
\end{itemize}

The point of this illustration is that the non-linear equation describing potential solitons is completely changed from that for an electron-positron plasma. The electron-positron plasma illustrated earlier has no turning points for the electrostatic potential taken on its own and so there are no soliton structures. In the case of the ion-electron plasma the corresponding field equation admits of two turning points and the analysis of the field equation in each domain has to be undertaken. It is this fact that the example has been used to illustrate. Every choice of distribution function must be investigated anew for the turning points and the corresponding analysis, i.\,e., the above case-by-case investigation, has to be undertaken in each regime.

If one chooses \emph{not} to have one of the components $A_j=0$ (with $j\in\{y,z\}$) then the electrostatic and electromagnetic components mix, representing a hybrid soliton. The point to make is that it is the symmetry breaking (for different distribution functions or the different masses of the charged particle species) that permits different types of soliton patterns, and that has been the purpose of the examples given here.

\section{The Relativistic Behavior}\label{rel}

In the situation where one cannot set $\gamma$ to unity, a considerably more complex behavior arises but, at the same time, one has the transverse $(v_y,v_z)$ particle velocities limited to $\pm c$ unlike in the non-relativistic situation where the corresponding limits are set as $\pm\infty$. This fundamental difference in behavior alters radically the soliton character.

Here one has
\be
\pd[f]x+\frac{em\gamma}{p_x}\,\pd[f]{p_x}\left[-\Ph'-\frac{1}{2me\gamma}{\left(\f\varpi\se-\frac{e}{c}\,\f A\se\right)^2}'\right]=0,
\ee
where
\be
\gamma=\sqrt{1+\frac{p_x^2}{(mc)^2}+\frac{1}{(mc)^2}\left(\f\varpi\se-\frac{e}{c}\,\f A\se\right)^2}
\ee
so that
\be
\pd[f]x+\frac{e}{mc^2}\,\pd[f]\gamma\left[-\Ph'+\frac{1}{2me\gamma}{\left(\f\varpi\se-\frac{e}{c}\,\f A\se\right)^2}'\right]=0.
\ee
The characteristic is given through
\be\label{eq:char}
\dd[\gamma]x=-\frac{e}{mc^2}\left[\Ph'-\frac{1}{2me\gamma}{\left(\f\varpi\se-\frac{e}{c}\,\f A\se\right)^2}'\right],
\ee
which has the basic structure
\be\label{eq:bstruct}
\dd[y]x=-a'(x)-\frac{1}{y}\,b'(x).
\ee
The relativistic form of the characteristic equation is not analytically tractable in general. However, in the case of interest where the influence of an electrostatic (electromagnetic) potential on an electromagnetic (electrostatic) soliton is sought one has either $\abs{a'}\ll\abs{b'}/\gamma$ or $\abs{a'}\gg\abs{b'}/\gamma$. In these two situations it is possible to derive approximate characteristics. Also, in the weakly relativistic limit where one can write $\gamma=1+\epsilon$ with only first order in $\epsilon$ terms being held in Eq.~\eqref{eq:char}, it is possible to perform a complete analytic investigation, thereby illuminating the transition between non-relativistic and relativistic limits. Consider this case first.

\subsection{Weakly Relativistic Behavior}\label{wrel}

Consider Eq.~\eqref{eq:bstruct} with $y=1+\epsilon$ and $\epsilon$ considered small, i.\,e., $\abs\epsilon\ll1$. Then to first order in $\epsilon$ one has
\be
\dd[\epsilon]x=-a'(x)-b'(x)+b'(x)\epsilon.
\ee
Then
\be
\dd x(\E{-b}\epsilon)=-\E{-b(x)}a'(x)+\dd x(\E{-b}),
\ee
with the solution
\be\label{eq:eps_char}
\epsilon=\epsilon_0\E{b(x)}+1-\E{b(x)}\int^x\df x'\;a'(x')\E{-b(x')},
\ee
where $\epsilon_0$ is a constant: the \emph{characteristic constant}. Here, for $a(x)$ one has
\bs
\begin{eqnarray}
&&a'(x)=\frac{e}{mc^2}\,\Ph'(x)\nonumber\\
\Longrightarrow\quad &&a(x)=\frac{e}{mc^2}\,\Ph(x)
\end{eqnarray}
and for $b(x)$, likewise,
\begin{eqnarray}
&&b'(x)=-\frac{1}{2(mc)^2}{\left(\f\varpi\se-\frac{e}{c}\,\f A\se\right)^2}'\nonumber\\
\Longrightarrow\quad &&b(x)=-\frac{1}{2(mc)^2}\left(\f\varpi\se-\frac{e}{c}\,\f A\se\right)^2.
\end{eqnarray}
\es
Then, by re-arranging the equation for $\gamma$, i.\,e.,
\be
\gamma\equiv1+\epsilon\equiv1+\frac{1}{2(mc)^2}\left[p_x^2+\left(\f\varpi\se-\frac{e}{c}\,\f A\se\right)^2\right]
\ee
an expression is provided for $p_x$ in terms of the characteristics $\epsilon_0$ and $\varpi_j$.

The distribution function $f=f(\epsilon_0,\varpi_y,\varpi_z)$ is an arbitrary function of its arguments. Now if $f(\epsilon_0,\varpi_y,\varpi_z)$ were to be taken as a function of $\epsilon_0$ only and not of $\varpi_j$ separately then, because $p_j$ is then a function of $(\varpi\se-e\f A\se/c)^2$ as is $\gamma$, it follows that transverse currents
\be
\f J\se\propto\int\frac{\df^3p}{\gamma}\,f(\epsilon_0)\left(\f \varpi\se-\frac{e}{c}\,\f A\se\right)
\ee
are identically zero, because the integrand is odd. Hence it is necessary and sufficient that $f(\epsilon_0,\varpi_y,\varpi_z)$ be a function of its arguments $\epsilon_0$ and $\varpi_y$ and/or $\varpi_z$ in order to have spontaneous symmetry breaking and so a transverse current.

Clearly, just as for the non-relativistic situation, different choices made for $f(\epsilon_0,\varpi_y,\varpi_z)$ determine the allowable soliton spatial structures. Even with the same choice of functional behavior in distribution functions for the non-relativistic and weakly relativistic situations one has different soliton structures. Consider, for example, the case of infinitely massive ions and mobile electrons with the electron distribution function being taken as
\be
\fl
f_e(\epsilon_0,\varpi_y,\varpi_z)=f_0\exp\!\left(-\frac{\epsilon_0}{\epsilon_\star}\right)\exp\!\left(-\frac{\varpi_y^2+\varpi_z^2}{\varpi_0^2}\right),\qquad f_0=\const.
\ee
Now in the weakly relativistic limit one can write the characteristic constant, $\epsilon_0$, as
\be\label{eq:eps0}
\fl
\epsilon_0=\E{-b(x)}\left\{-1+\E{b(x)}\int^x\df x'\;a'(x')\E{-b(x')}+\frac{1}{2(mc)^2}\left[p_x^2+\left(\varpi\se-\frac{e}{c}\,A\se\right)^2\right]\right\},
\ee
thereby expressing $p_x$ in terms of the characteristics, i.\,e, $\epsilon_0$ as defined in Eq.~\eqref{eq:eps_char}.

Now consider the transverse current integral
\bs
\begin{eqnarray}
\f J\se&=&e\int\df^3p\;\f v\se f(\epsilon_0,\varpi_y,\varpi_z)\\
&\equiv&\frac{e}{m}\int\frac{\df^3p}{\gamma}\left(\f\varpi\se-\frac{e}{c}\,\f A\se\right)f(\epsilon_0,\varpi_y,\varpi_z).
\end{eqnarray}
\es
Using $1/\gamma=1/(1+\epsilon)\approx1-\epsilon\approx\exp(-\epsilon)$, one can then write, in the weakly relativistic limit,
\begin{eqnarray}
\fl
\f J\se&=&\frac{e}{m}\int\df p_x\,\df\varpi_y\,\df\varpi_z\,\left(\f\varpi\se-\frac{e}{c}\,\f A\se\right)f_0\nonumber\\
\fl
&\times&\exp\!\left(-\frac{\f\varpi\se^2}{\varpi_0^2}\right)\exp\!\left\{-\frac{\epsilon_0}{\epsilon_\star}-\frac{1}{2(mc)^2}\left[p_x^2+\left(\f\varpi\se-\frac{e}{c}\,\f A\se\right)^2\right]\right\}.
\end{eqnarray}
With $\epsilon_0$ being expressed in terms of $p_x^2$ and $(\f\varpi\se-e\f A\se/c)^2$ through Eq.~\eqref{eq:eps0} the integral over $p_x$ can be performed immediately yielding
\begin{eqnarray}
\fl
\f J\se&=&\frac{e}{c}\sqrt{2\pi}\left(1+\frac{\E{-b(x)}}{\epsilon_\star}\right)^{1/2}\exp\!\left[\frac{\E{-b(x)}}{\epsilon_\star}\left(-1+\E{b(x)}\int^x\df x'\;a'(x')\E{-b(x')}\right)\right]f_0\nonumber\\
\fl
&\times&\int\df\varpi_y\,\df\varpi_z\,\left(\f\varpi\se-\frac{e}{c}\,\f A\se\right)\exp\!\left(-\frac{\f\varpi\se^2}{\varpi_0^2}\right)\nonumber\\
\fl
&\times&\exp\!\left[-\frac{1}{2(mc)^2}\left(\f\varpi\se-\frac{e}{c}\,\f A\se\right)^2\left(1+\frac{\E{-b(x)}}{\epsilon_\star}\right)\right].
\end{eqnarray}
The double integral over $\varpi_y$ and $\varpi_z$ can be done in closed form so that one can write
\begin{eqnarray}
\fl
\f J\se&=&\sqrt{2\pi^3}\,\frac{\varpi_0^2}{\left(1+\beta\varpi_0^2\right)^2}\,\exp\!\left\{-\frac{\E{-b(x)}}{\epsilon_\star}\left[-1+\E{b(x)}\int^x\df x'\;a'(x')\E{-b(x')}\right]\right\}f_0\nonumber\\
\fl
&\times&\exp\!\left[-\frac{\E2}{c^2}\,\frac{\beta}{1+\beta\omega_0^2}\left(A_y^2+A_z^2\right)\right]\f A\se
\end{eqnarray}

Even in this seemingly simple extension of the non-relativistic results to the weakly relativistic situation one is challenged by an exceedingly non-linear set of equations for the field components. How many solutions the equations admit, how the elctrostatic and electromagnetic components are coupled, and how the solution(s) structure depends on the various parameters remain analytically intractable but are likely best addressed by using numerical procedures.

The point being illustrated here is that, despite the characteristics being available in closed analytic form, the non-linear complexities of the current distribution as functions of the electrostatic and electromagnetic potentials are less than inviting. And for each choice of distribution function similarly complex results obtain. It would seem that only numerical procedures can help.

\subsection{Fully relativistic behavior}\label{frel}

Because the characteristic Eq.~\eqref{eq:char} is not solvable analytically in general, two limiting cases will be considered in what follows.

\subsubsection{The case $\abs{a'}\ll\abs{b'}/\gamma$}\label{case1}

Here write the characteristic equation as
\be
\gamma\,\dd[\gamma]x+b'=-a'\gamma.
\ee
Then
\be
\dd x\left(\frac{1}{2}\,\gamma^2+b\right)=-a'\gamma
\ee
so that
\be
\frac{1}{2}\,\gamma^2+b=-\int^x\df\xi\;a'\gamma+\lambda=-a\gamma+\int^x\df\xi\;a\gamma'+\lambda.
\ee
In the integral on the right-hand side set, on evaluating $\gamma^2/2+b\approx\lambda$,
\be
\gamma'=-\frac{b'}{\sqrt{2\left(\lambda-b\right)}}
\ee
to lowest order. Then
\be
\frac{1}{2}\,\gamma^2+b+a\gamma+\int\df\xi\;\frac{ab'}{\sqrt{2\left(\lambda-b\right)}}=\lambda
\ee
and $f$ is a function solely of $\lambda$, $\varpi_y$, and $\varpi_z$.

\subsubsection{The case $\abs{a'}\gg\abs{b'}/\gamma$}\label{case2}

Here write
\be
\pd[\gamma]x+a'=-\frac{b'}{\gamma}
\ee
so that
\be
\dd x\left(\gamma+a\right)=-\frac{b'}{\gamma}
\ee
with, then,
\be\label{eq:gpa}
\gamma+a=-\int^x\df\xi\;\frac{b'}{\gamma}+\lambda.
\ee
In the integration on the right-hand side set $\gamma=\lambda-a$, (i.\,e., solve for $\gamma$ by neglecting the integral itself) to obtain
\be
\gamma+a+\int^x\df\xi\;\frac{b'}{\lambda-a}=\lambda
\ee
to lowest order, which defines the characteristic to order $\abs{b'/(a'\gamma)}$ as required. Then $f$ is a function solely of $\lambda$, $\varpi_y$, and---in contrast to Sec.~\ref{case1}---of $a$. Because $\abs{a'}$ is large, a direct dependence on $a$ is the most useful choice in this case.\footnote{Because $f$ is an arbitrary (but positive) function of its arguments one can choose many different functional forms to illuminate particular points, which is what is done throughout the paper.}

Note that the characteristic Eq.~\eqref{eq:char} is symmetric in $(\f\varpi\se-e\f A\se/c)^2$ so that if the distribution function is \emph{not} explicitly dependent on $\varpi_y,\varpi_z$ but only on the constant arising from the characteristic Eq.~\eqref{eq:char}, then $\f A\se=0$ by symmetry arguments. Thus, just as for the non-relativistic limit: only when the distribution function depends \emph{explicitly} on $\varpi_y$ and/or $\varpi_z$, in addition to the constant arising from the characteristic equation is there the possibility of a self-consistent electromagnetic soliton because of the asymmetry of the current integrals with respect to $\f\varpi\se-e\f A\se/c$. However, the functional form of such solitons (and also the modifications brought about by coupling of the electrostatic and electromagnetic fields) is considerably different than their non-relativistic counterparts due to the relativistic limitation that particle speeds must be less than $c$.

To illustrate this basic point consider again an electron-positron plasma in which the two distribution functions are identical. To make the comparison as close as possible between the relativistic and non-relativistic situations consider the electron and positron distribution functions [as functions of $\varpi_y,\varpi_z$, and $\lambda$, the characteristic from Eq.~\eqref{eq:gpa}] to be given by
\be
f=f_0\,\exp\!\left[-\frac{\lambda}{\lambda_0}-\frac{\varpi_y^2+\varpi_z^2}{\varpi_0^2}\right]
\ee
with $f_0,\lambda_0$, and $\varpi_0$ constants.

Then consider the current integral
\be
J_y=\int\frac{\df^3p}{\gamma}\left(\varpi_y-\frac{e}{c}\,A_y\right)f.
\ee

While the general relativistic equation is solvable analytically only in the non-relativistic and weakly relativistic situations, it has the property that it depends only on the combination $(\f\varpi\se-e\f A\se/c)^2$. Thus the characteristic constant for the equation (say, \La) is also a function solely of $(\f\varpi\se-e\f A\se/c)^2$.

Hence, for distribution functions that are functions in the form $f(\La,\varpi_y,\varpi_z)$ it follows that if $f$ is chosen to be a function solely of \La\ and not of $\varpi_y,\varpi_z$ then all transverse current components are precisely zero. Under such conditions there are no electromagnetic soliton solutions. This aspect has already been seen in the non-relativistic (Sec.~\ref{nonrel}) and weakly relativistic (Sec.~\ref{wrel}) cases, and is now of general validity. Thus $f$ must be a function of $\varpi_y$ and/or $\varpi_z$ as well as \La\ in order to obtain an electromagnetic soliton. Despite the fact one cannot solve the characteristic equation in closed form for a fully relativistic plasma as shown in \ref{r_transf}, one can obtain accurate approximate solutions when the electrostatic field is either small or large compared to the Lorentz force per unit charge. In both cases it is then possible to express $\gamma$ in terms of the characteristic constant, as also detailed in \ref{r_transf}.

The existence and structure of any solitons (electromagnetic and/or electrostatic) then depends on the choices made for the particle distribution functions, as also exhibited in detail for the non-relativistic and weakly relativistic solutions.

\section{Summary and Discussion}\label{summ}

While the linear Weibel instability has been thoroughly investigated, regrettably the same cannot be said of the non-linear behavior including the coupling of electromagnetic and electrostatic effects. The exploration of the non-linear aspects given here has uncovered a variety of effects that are germane to future investigation for both non-relativistic and relativistic plasmas. While in both the non-relativistic and the relativistic case, the classical constants of motion are total energy and generalized momentum, the problem is to obtain a \emph{closed form} expression for $\gamma$ (or, in the notation used here, $p_x$) in terms of these constants. Because of the coupled effects of the magnetic fields and the electrostatic fields such appears not to be an easily tractible problem as shown in text.

It has been shown previously that Weibel isolated modes (which, subsequently, can develop soliton modes) are retained in analytical calculations even if one allows for ``classic'' extended unstable wavenumber ranges. Because such structures develop only in asymmetric plasmas, they serve as an indicator for the asymmetry of the particle distribution function. Therefore, because precisely symmetric plasmas are difficult to achieve in nature, isolated Weibel modes will be ubiquitous, as shown recently \citep{tau07:wea}. Furthermore, even if the unstable wave modes are allowed to have a ``weak'' propagating component, the isolated Weibel modes are still generated. Hence, soliton structures should \emph{always} be taken into consideration when investigating: (i) instabilities in (relativistic) plasmas in general; (ii) non-linear behavior of the resulting unstable modes; (iii) particle radiation patterns due to scattering in such modes.

Perhaps the most significant theme is that the occurrence of a non-linear Weibel-like soliton requires that the distribution functions be dependent on all three of the characteristic constants. Without such a dependence (and in particular with no explicit dependence on characteristic constants perpendicular to the spatial variation direction of the soliton) then there is no electromagnetic current and so no soliton. This point was demonstrated for non-relativistic, weakly relativistic, and fully relativistic plasma situations.

Even then, the functional behavior of the distribution functions on the three characteristic variables was shown, by explicit examples, to play a fundamental role in determining the structure of the non-linear equations for the coupled electromagnetic and electrostatic fields. Cases were given where no soliton was possible, where solitons existed only for decoupled electromagnetic field with no electrostatic component, and where changes in the distribution functions altered the non-linear field equations so markedly that each situation had to be considered anew. The characteristic constants could be written down in closed form for the non-relativistic and weakly relativistic situations, and the constants could be approximated in the fully relativistic plasma situation for weak electrostatic (electromagnetic) effects on an electromagnetic (electrostatic) field.

Nevertheless the complexity of the resulting non-linear field equations is daunting. Except for simple situations it has so far not proven possible to solve such non-linear equations in either particular cases or the general case for chosen distribution functions. One suspects that only numerical procedures will allow deeper insight into the classes of functional behavior for distribution functions that allow solitons, for the spatial structure of such solitons, and for the relative contributions of the electrostatic and electromagnetic fields to any such solitons. Future work should attempt to investigate the modifications of the radiation pattern due to particle scattering in such soliton structures. In doing so, the question can be explored if and to what extent the radiation spectrum in relativistic outflows deviates from pure synchrotron radiation.

\ack
{\it We thank the anonymous referee for scrutizining our manuscript ever so thoroughly. This work was partially supported by the Deutsche Forschungsgemeinschaft (DFG) through grant No.\ \mbox{Schl~201/17-1}, Sonderforschungsbereich 591, and by the German Academy of Natural Scientists Leopoldina Fellowship Programme through grant LDPS 2009-14 funded by the Federal Ministry of Education and Research (BMBF).}

\appendix

\section{Non-Relativistic Integral Transformations}\label{nr_transf}

Transformation of the non-relativistic charge and current density integrals proceeds as follows. In terms of the variables
\begin{eqnarray}
w&=&p_x^2+2m_\alpha e_\alpha\Ph+\left(\varpi_y-\frac{e_\alpha}{c}\,A_y\right)^2+\left(\varpi_z-\frac{e_\alpha}{c}\,A_z\right)\label{eq:w}\\
\varpi_{(y,z)}&=&p_{(y,z)}+\frac{e_\alpha}{c}\,A_{(y,z)}
\end{eqnarray}
one has
\be
I_Q\equiv\int\df^3p\;f(w,\varpi_y,\varpi_z)
\ee
and
\be
\f I_J=\int\df^3p\,\left(\f\varpi\se-\frac{e_\alpha}{c}\,\f A\se\right)f(w,\varpi_y,\varpi_z), \label{eq:IJ}
\ee
where $\f\varpi\se=(0,\varpi_y,\varpi_z)$ and $\f A\se=(0,A_y,A_z)$. The volume element $\df^3p$ is written in terms of $\df w\,\df\varpi_y\,\df\varpi_z$ using a conventional Jacobian procedure, yielding
\be
\df^3p=\frac{\df w\,\df\varpi_y\,\df\varpi_z}{2\sqrt{w-w_\star}}
\ee
with $w_\star=2e_\alpha m_\alpha\Ph+\left(\f\varpi\se-e_\alpha\f A\se/c\right)^2$.

Then, by taking into account that according to the definition Eq.~\eqref{eq:w}, $w$ is symmetric in $\pm p_x$ and that hence one has two integration regimes $w_\star\leqslant w\leq\infty$ for the integration range $-\infty<p_x<\infty$, one obtains
\be
\fl
I_Q=\int_{w_\star}^\infty\df w\uint\df\varpi_y\uint\df\varpi_z\,\left(w-w_\star\right)^{-1/2}f(w,\varpi_y,\varpi_z),
\ee
By defining $\xi=w-w_\star$, this integral can be expressed as
\be\label{eq:IQ}
\fl
I_Q=-2\int_0^\infty\df\xi\uint\df\varpi_y\uint\df\varpi_z\;\sqrt\xi\,\pd\xi\,f\!\left(w_\star+\xi,\varpi_y,\varpi_z\right).
\ee
Likewise, the current integral from Eq.~\eqref{eq:IJ} takes on the form
\begin{eqnarray}
\f I_J&=&-2\int_0^\infty\df\xi\uint\df\varpi_y\uint\df\varpi_z\;\sqrt\xi\nonumber\\
&\times&\left(\f\varpi\se-\frac{e_\alpha}{c}\,\f A\se\right)\pd\xi\,f\!\left(w_\star+\xi,\varpi_y,\varpi_z\right). \label{eq:IJ2}
\end{eqnarray}
Eqs.~\eqref{eq:IQ} and \eqref{eq:IJ2} are used in text.

\section{Relativistic Integral Transformations}\label{r_transf}

As for the nonrelativistic situation, transformations in the relativistic situation rely upon the connection between the momentum variables and the canonical variables $\lambda,\varpi_y$, and $\varpi_z$. A simple Jacobian transformation shows that
\be
\df^3p=\frac{1}{2}\,\frac{(mc)^2}{\abs{p_x}}\abs{\pd[\gamma^2]\lambda}\df\lambda\,\df\varpi_y\,\df\varpi_z,
\ee
where one has
\be\label{eq:Btmp1}
\pd[\gamma]x=-a'(x)-\frac{1}{\gamma}\,b'(x)
\ee
with
\be\label{eq:BgaD}
\frac{1}{2}\,\gamma^2+b(x)+\int_0^x\df x'\;a'(x')\gamma(x')=\lambda.
\ee

From the derivative of Eq.~\eqref{eq:BgaD} with respect to $\lambda$ one obtains
\be
\frac{1}{2}\,\pd[\gamma^2]\lambda=1-\frac{1}{2}\int_0^x\df x'\;\frac{a'(x')}{\gamma(x')}\,\pd[\gamma(x')^2]\lambda
\ee
or, from Eq.~\eqref{eq:Btmp1}
\be
\pd x\left(\pd[\gamma^2]\lambda\right)=-\frac{a'(x)}{\gamma}\,\pd[\gamma^2]\lambda
\ee
Then, generally, one can write
\be\label{eq:Btmp2}
\pd[\gamma^2]\lambda=\exp\left[-\int_0^x\df x'\;\frac{a'(x')}{\gamma(x')}\right]>0.
\ee
Note also that
\be\label{eq:Btmp3}
\pd[\gamma]\lambda=\exp\left[\int_0^x\df x'\;\frac{b'(x')}{\gamma(x')^2}\right]>0.
\ee
The representations from Eqs.~\eqref{eq:Btmp2} and \eqref{eq:Btmp3} are particularly useful when discussing the limiting cases $\abs{b'}\gg\abs{a'\gamma}$ and $\abs{b'}\ll\abs{a'\gamma}$.

Because $\gamma\geqslant1$ always it follows that
\be\label{eq:La}
\lambda\geqslant \frac{1}{2}+b'(x)+\int_0^x\df x'\;a'(x')\gamma(x')\equiv\La(x)
\ee
for \emph{all} $x$ so that $\La(x)$ must have a maximum value.

Now Eq.~\eqref{eq:Btmp1} is not solvable analytically in closed form for arbitrary $a(x)$ and $b(x)$. Thus it is appropriate to investigate the limiting cases of $\abs{b'}\gg\abs{a'\gamma}$ and $\abs{b'}\ll\abs{a'\gamma}$. Consider each in turn.

\subsection{Case 1. $\abs{b'}\gg\abs{a'\gamma}$}

Here terms to leading order in $a'$ will be retained because they represent the small order electrostatic corrections to an otherwise pure electromagnetic situation. One can also write
\be\label{eq:gamma}
\gamma+a(x)+\int_0^x\df x'\;\frac{b'(x')}{\gamma(x')}=\nu,
\ee
which, with $\gamma\geqslant1$ everywhere, implies the constant $\nu$ satisfies $\nu\geqslant\nu_m$, where $\nu_m$ is the maximum value of the left-hand side of Eq.~\eqref{eq:gamma}.

A connection between $\lambda$ and $\nu$ is obtained simply by inserting $\gamma$ from Eq.~\eqref{eq:gamma} into Eq.~\eqref{eq:La} so that of Eq.~\eqref{eq:BgaD}
\be
\fl
\frac{1}{2}\left[\nu-a(x)-\int_0^x\df x'\;\frac{b'(x')}{\gamma(x')}\right]^2+b(x)+\int_0^x\df x'\;a'(x')\gamma(x')=\lambda.
\ee
Consider then the volume element
\begin{eqnarray}
\df^3p&=&\frac{1}{2}\,\frac{(mc)^2}{\abs{p_x}}\abs{\pd[\gamma^2]\lambda}\df\lambda\,\df\varpi_y\,\df\varpi_z \nonumber\\
&=&\frac{(mc)^2}{\abs{p_x}}\,\exp\!\left[-\int_0^x\df x'\;\frac{a'(x')}{\gamma(x')}\right]\df\lambda\,\df\varpi_y\,\df\varpi_z.
\end{eqnarray}
Write
\begin{eqnarray}
\frac{\abs{p_x}}{mc}&=&\left[\gamma^2-1-\frac{1}{(mc)^2}\left(\f\varpi\se-\frac{e}{c}\,\f A\se\right)^2\right]^{1/2} \nonumber\\
&=&\Biggl[2\left(\lambda-b(x)-\int_0^x\df x'\;a'(x')\gamma(x')\right)\nonumber\\
&-&1-\frac{1}{(mc)^2}\left(\f\varpi\se-\frac{e}{c}\,\f A\se\right)^2\Biggr]^{1/2}.
\end{eqnarray}
Interest centers on the expansion of the volume element to lowest order in $a(x)$. Then
\be
\int_0^x\df x'\;\frac{a'(x')}{\gamma(x')}\simeq\int_0^x\df x'\;a'(x')\left/\sqrt{2\left[\lambda-b(x')\right]}\right.
\ee
and
\be
\int_0^x\df x'\;a'(x')\gamma(x')\simeq\int_0^x\df x'\;a'(x')\sqrt{2\left[\lambda-b(x')\right]}
\ee
so that
\begin{eqnarray}
\df^3p&\simeq&(mc)\,\df\lambda\,\df\varpi_y\,\df\varpi_z\left(1-\int_0^x\df x'\frac{a'(x')}{\sqrt{2\left[\lambda-b(x')\right]}}\right) \nonumber\\
&\times&\Biggl\{2\left[\lambda-b(x)\right]-1-\frac{1}{(mc)^2}\left(\f\varpi\se-\frac{e}{c}\,\f A\se\right)^2\Biggr\}^{-1/2} \nonumber\\
&\times&\Biggl\{1+\int_0^x\df x'\;a'(x')\sqrt{2\left[\lambda-b(x')\right]}\nonumber\\
&\times&\left[2\left[\lambda-b(x)\right]-1-\frac{1}{(mc)^2}\left(\f\varpi\se-\frac{e}{c}\,\f A\se\right)^2\right]^{-1}\Biggr\}.
\end{eqnarray}
Collecting terms to leading order in $a$ yields
\begin{eqnarray}\label{eq:d3pA}
\fl
\df^3p&\simeq&(mc)\,\df\lambda\,\df\varpi_y\,\df\varpi_z\left\{2\left[\lambda-b(x)\right]-1-\frac{1}{(mc)^2}\left(\f\varpi\se-\frac{e}{c}\,\f A\se\right)^2\right\}^{-1/2} \nonumber\\
\fl
&\times&\Biggl\{1-\int_0^x\df x'\;\frac{a'(x')}{\sqrt{2\left[\lambda-b(x')\right]}}+\int_0^x\df x'\;a'(x')\sqrt{2\left[\lambda-b(x')\right]} \nonumber\\
\fl
&\times&\left[2\left[\lambda-b(x)\right]-1-\frac{1}{(mc)^2}\left(\f\varpi\se-\frac{e}{c}\,\f A\se\right)^2\right]^{-1}\Biggr\}.
\end{eqnarray}


\subsection{Case 2. $\abs{b'}\ll\abs{a'\gamma}$}

In this situation, representing an electrostatic field dominating over electromagnetic contributions one retains terms to leading order in $b$. Then one uses again
\begin{equation*}
\pd[\gamma]\lambda=\exp\left[\int_0^x\df x'\;\frac{b'(x')}{\gamma(x')^2}\right]
\end{equation*}
and
\be
\gamma+\int_0^x\df x'\;\frac{b'(x')}{\gamma(x')}=\nu-a(x)
\ee
so that
\be\label{eq:Btmp4}
\pd[\gamma]\nu=\exp\left[\int_0^x\df x'\;\frac{b'(x')}{\gamma(x')^2}\right].
\ee
Hence
\be
\df^3p=\frac{(mc)^2}{\abs{p_x}}\,\gamma\abs{\pd[\gamma]\nu}\,\df\nu\,\df\varpi_y\,\df\varpi_z.
\ee

To leading terms in $b$ one can replace $\gamma(x')^2$ in Eq.~\eqref{eq:Btmp4} by $[\nu-a(x')]^2$; equally one can write
\be
\gamma(x)=\left[\nu-a(x)-\int_0^x\df x'\;\frac{b'(x')}{\nu-a(x')}\right].
\ee
Then noting that
\be
\abs{p_x}=mc\left[\gamma^2-1-\frac{1}{(mc)^2}\left(\f\varpi\se-\frac{e}{c}\,\f A\se\right)^2\right]^{1/2}
\ee
one has
\begin{eqnarray}\label{eq:Btmp5}
\fl
\df^3p&=&(mc)\left[\nu-a(x)-\int_0^x\df x'\;\frac{b'(x')}{\nu-a(x')}\right]\left[1+\int_0^x\df x'\;\frac{b'(x')}{\left[\nu-a(x')\right]^2}\right] \nonumber\\
\fl
&\times&\Biggl\{\left[\nu-a(x)-\int_0^x\df x'\;\frac{b'(x')}{\nu-a(x')}\right]^2\nonumber\\
\fl
&-&1-\frac{1}{(mc)^2}\left(\f\varpi\se-\frac{e}{c}\,\f A\se\right)^2\Biggr\}^{-1/2}\df\nu\,\df\varpi_y\,\df\varpi_z.
\end{eqnarray}
Expanding the right-hand side of Eq.~\eqref{eq:Btmp5} to leading terms in $b$ yields
\begin{eqnarray}\label{eq:d3pB}
\fl
\df^3p&=&(mc)\,\df\nu\,\df\varpi_y\,\df\varpi_z\left[\left(\nu-a(x)\right)^2-1-\frac{1}{(mc)^2}\left(\f\varpi\se-\frac{e}{c}\,\f A\se\right)^2\right]^{-1/2} \nonumber\\
\fl
&\times&\left[\nu-a(x)\right]\Biggl\{1-\left[\nu-a(x)\right]^{-1}\int_0^x\df x'\;\frac{b'(x')}{\nu-a(x')}\nonumber\\
\fl
&+&\int_0^x\df x'\;\frac{b'(x')}{\left[\nu-a(x')\right]^2}+\left[\nu-a(x)\right]\int_0^x\df x'\frac{b'(x')}{\nu-a(x')}\nonumber\\
\fl
&\times&\left[\left(\nu-a(x)\right)^2-1-\frac{1}{(mc)^2}\left(\f\varpi\se-\frac{e}{c}\,\f A\se\right)^2\right]^{-1}\Biggr\}.
\end{eqnarray}


In addition to the volume element $\df^3p$ transformed to the canonical variables, in the integrals for transverse current contributions one also requires the elements $\df^3p/\gamma$. These elements follow directly from the above discussion because in the case $\abs{b'}\ll\abs{a'/\gamma}$ one has
\be
\gamma\simeq\nu-a(x)-\int_0^x\df x'\;\frac{b'(x')}{\nu-a(x')},
\ee
while in the case $\abs{b'}\gg\abs{a'/\gamma}$ one has
\begin{eqnarray}
\gamma&=&\left\{2\left[\lambda-b(x)-\int_0^x\df x'\;a'(x')\gamma(x')\right]\right\}^{1/2} \nonumber\\
&\simeq&\left\{2\left[\lambda-b(x)-\int_0^x\df x'\;a'(x')\sqrt{2\left[\lambda-b(x')\right]}\right]\right\}^{1/2}.
\end{eqnarray}
Then
\be\label{eq:tmp6}
\frac{1}{\gamma}=\frac{1}{\nu-a(x)}\left[1+\frac{1}{\nu-a(x)}\int_0^x\df x'\;\frac{b'(x')}{\nu-a(x')}\right]
\ee
in the case $\abs{b'}\ll\abs{a'/\gamma}$, while when $\abs{b'}\gg\abs{a'/\gamma}$ one has
\be
\fl
\frac{1}{\gamma}\simeq\frac{1}{\sqrt{2\left[\lambda-b(x)\right]}}\left\{1+\frac{1}{2}\left[\lambda-b(x)\right]^{-1}\int_0^x\df x'\;a'(x')\sqrt{2\left[\lambda-b(x')\right]}\right\}
\ee
so that one has immediate expressions by combining Eqs.~\eqref{eq:tmp6} with \eqref{eq:d3pB} (in the case $\abs{b'}\gg\abs{a'\gamma}$).


\bibliography{../../tautz,../../book,../../article}
\bibliographystyle{jphysicsB}

\end{document}